\documentclass[showpacs,preprint]{revtex4}

\def\be{\begin{equation}}
\def\ee{\end{equation}}
\def\beq{\begin{eqnarray}}
\def\eeq{\end{eqnarray}}
\def\n{\nonumber}

\usepackage{graphicx}
\usepackage{dcolumn}
\usepackage{amsmath}

\begin{document}


\title{Expanding, shearing and accelerating isotropic plane symmetric universe  with conformal Kasner geometry}



\author{Sudan Hansraj$^a$}
\author{Megandhren Govender$^{a,b}$}
\author{Narenee Mewalal$^{a,b}$}

\affiliation{$^a$Astrophysics and Cosmology Research Unit, School of Mathematics, University of KwaZulu Natal, Private Bag X54001,
              Durban 4000, South Africa.
              Tel.: +27 31 260 3000
              Fax: +27 31 260 2632 email: hansrajs@ukzn.ac.za }
              \affiliation{$^b$Department of Mathematics, Durban University of Technology, Durban, 4000, South Africa.}


\date{\today}

\begin{abstract}
We construct a model of a universe filled with a perfect fluid with isotropic particle pressure. The anisotropic plane symmetric Kasner spacetime is used as a seed metric and through a conformal mapping a perfect fluid is generated. The  model is inhomogeneous,  irrotational, shearing and accelerating. For negative time, the universe is expanding while for positive time it is collapsing. With the aid of graphical three dimensional plots it is established that the density and pressure hypersurfaces are smooth and singularity free. Additionally the sound-speed index is computed and the fluid obeys causality criterion.

\end{abstract}

\pacs{04.20.Jb, 04.40.Nr, 04.70.Bw}

\maketitle


\section{Introduction}

Following the recent detection of gravitational waves by LIGO \cite{abbott}, interest in non-spherically symmetric solutions of Einstein's equations has increased.  It is well known through the Jebsen--Birkhoff theorem  that spherically symmetric pulsating objects do not generate gravitational waves therefore other symmetry options should be studied. In light of this, we consider plane symmetric spacetimes and endeavour to construct exact solutions that may be utilised to model perfect fluids. Such solutions may be both of cosmological or astrophysical interest depending on the existence on a hypersurface of vanishing pressure.

Plane symmetric spacetimes  admit a three-parameter group of motions in the Euclidean plane. Taub \cite{taub} demonstrated that spacetimes which possess plane symmetry have properties similar to those of spherically symmetric spacetimes. At its present stage of evolution the universe, as described by the Friedman-Robertson-Walker model, is spherically symmetric with matter distribution that is homogenous and isotropic. However, these assumptions of spherical symmetry and isotropy were not necessarily valid at the time of early evolution. Hence many authors consider plane symmetry with a perfect fluid source in an attempt to solve the Einstein field equations.

Some earlier studies on plane symmetric perfect fluids, in various contexts, include those conducted by Roy and Narain \cite{roynar}, Tikekar \cite{tikekar}, Collins \cite{collins} and Davidson \cite{davidson}. Davis and Ray \cite{davis} solved the Einstein--Dirac field equations for a ghost neutrino field and produced a model with shear, rotation and expansion. Sharif \cite{sharif} classified plane symmetric spacetimes according to matter collineations for both a degenerate and non-degenerate energy momentum tensor. Gr{\o}n and Soleng \cite{gron} analysed static plane symmetric fields conformally related to a massless scalar field.  Coley and Czapor \cite{coley} examined the inheriting conformal Killing vector properties of plane symmetric perfect fluid spacetimes.  Recently Zhang and Noh \cite{zhangnoh} found an $n$-dimensional ($n \geq 4$) plane symmetric solution for which the source was a perfect fluid. Anguige \cite{anguige} proved the existence of a class of plane symmetric perfect fluid cosmologies that become Kasner-like on approach to the initial singularity. In this paper we use the Kasner metric, together with a conformal transformation to generate and solve the Einstein field equations for a plane symmetric perfect fluid.

Locating exact solutions for various isotropic matter configurations has proved to be a nontrivial exercise on account of the severe nonlinearity of the associated field equations. Matter-free (vacuum) solutions are found more easily as the vanishing of the Ricci tensor is required. In the case of matter, it is well known that if anisotropic solutions are sought, then practically any metric may solve the Einstein field equations as there are more unknowns than there are governing  equations. When the condition of pressure isotropy is introduced the mathematical complexity increases. Nevertheless there exists an abundance of exact solutions for spherically symmetric spacetimes for isotropic perfect fluids. For example, see Delgaty and Lake  \cite{delgaty} and the references therein. This is not the case for plane symmetric distributions. The exterior of such systems are known to be described by the Kasner \cite{kasner} metric or the Taub \cite{taub} metric. In spherical symmetry, Birkhoff's theorem guarantees the existence of a unique exterior metric - the Schwarzschild solution, which is independent of whether the interior is static or nonstatic. There is no such counterpart for plane symmetry.

A mathematical technique to generate new exact solutions of Einstein's equations is through the use  of conformal transformations. In this approach, a known exact solution, such as a vacuum metric, serves as a seed solution. By performing a conformal transformation of such a metric and involving a nonzero energy momentum tensor, solutions describing the gravitational field of a perfect fluid may be constructed. For example, starting with the Schwarzschild \cite{schwarz} exterior geometry, Hansraj \cite{hans2} was able to generate the corresponding perfect fluid metric through a conformal mapping. Although the conformal factor was assumed to be time-dependent at the outset, it was proved that  all perfect fluids conformally  related to the  Schwarzschild exterior were necessarily static.  In general, conformal transformations do not preserve the symmetry  on account of a difference with the Killing vectors. A similar programme was attempted using the plane symmetric Taub \cite{hans3} metric and again, it was found that the emerging spacetime was necessarily static. Rich classes of solutions with physically palatable properties were reported in Hansraj {\it {et al}} \cite{hansgovmewa,hans1} with the help of Lie group analysis.
It is therefore interesting to ask if we cannot generate nonstatic solutions through a conformal transformation. A useful candidate to assist answering this question is the Kasner \cite{kasner} spacetime that itself is time dependent. The Kasner metric is known to be anisotropic and this feature has proved useful in developing models of the early universe where anisotropy may have been evident. In contrast it is known that the universe today is understood to be isotropic.  In this study we analyse whether a nonstatic solution with isotropic particle pressure can be established through the conformal transformations method on an anisotropic spacetime. The answer turns out to be in the affirmative.

The Kasner model presents a solution that is plane-symmetric as well as spatially homogenous and is used in modeling the early-time evolution of the universe \cite{anni}.  The role of anisotropy within the realm of cosmology has been demonstrated in various models of the universe ranging from inhomogeneous cosmologies, Lemaitre-Tolman-Bondi cosmological models, dark energy inspired models, phantom cosmologies as well as braneworld models. On the observational front, it is widely accepted that anisotropy played a fundamental role in the early expansion of the universe leading to structure formation, anisotropy in the CMBR, baryogenesis as well as the acceleration of the universe. For an exposition to these topics see the works of \cite{takeo,fuku, saaidi}  and the references therein.

We briefly survey some important  investigative works that have been conducted on the Kasner spacetime.
 The original form written by Kasner in 1921 was presented with a positive signature as
\[
ds^2 = x_1^{2a_1}dx_1^2 + x_1^{2a_2}dx_2^2 + x_1^{2a_3}dx_3^2 + x_1^{2a_4}dx_4^2
\]
in Wainwright and Krasinski \cite{wainkras}.
However, this form of the metric has been almost forgotten in favour of other forms. For example, see the two forms presented by Bedran {\it {et al}} \cite{bedran}.    Taub \cite{taub} derived the Kasner metric in its current, more familiar form
\[
ds^2 = - dt^2 + t^{2p_1}dx_1^2 + t^{2p_2}dx_2^2 + t^{2p_3}dx_3^2
\]
The Bianchi 1 model, which is obtained from the Kasner metric, has led to a large class of inhomogenous solutions as well as large families of Bianchi models. Recently, Govender and Thirukkanesh \cite{gov1} studied relaxational effects in a Bianchi I cosmological model within the framework of extended irreversible thermodynamics. They showed that relaxational effects lead to higher temperature of the cosmological fluid. The Belinsky-Khalatnikov-Lifshitz (BKL) \cite{bkl} model utilised the Kasner solution to describe the oscillatory nature of the universe around a gravitational singularity. The Mixmaster universe \cite{mix}  was a cosmological model proposed by Charles Misner in an attempt to understand the dynamics of the early universe. With just these few examples we note that the Kasner metric forms the background for the generation of a general family of cosmological solutions.

The structure of this paper is as follows: In section one we provide an overview of conformal geometry and its impact on the geometric and dynamical variables of a gravitating system. We then proceed to use the Kasner metric as a seed to generate the Einstein field equations for plane symmetric perfect fluids in section two. In section three an exact solution to the system of partial differential equations is obtained and is then studied for physical plausibility. Specifically we check the behavior of the pressure and energy density with the help of three dimensional plots. We conclude with a discussion of our findings.

\section{Conformal Riemannian Geometry}

For a detailed exposition of conformal geometry relevant to this study, the reader is referred to the work of Hansraj {\it {et al}} \cite{hansgovmewa}. In brief, suppose that we are given a spacetime $(M,{\bf g})$ with line element $
ds^2 = g_{ab} dx^adx^b
$
and a related spacetime $( M, {\bar{\bf g}})$ with the line element
$
d{\bar s}^2 = {\bar g}_{ab} dx^adx^b\;. \label{1d}
$
Then the above two line elements are said to be conformally related if
$$
{\bar g}_{ab} = e^{2U} g_{ab}\;\;{\rm and}\;\;
{\bar g}^{ab} = e^{-2U} g^{ab}\;, $$
where $U(x^c)$ is a nonzero, real--valued function of the coordinates on
$M$.
The metric connection, Riemann
curvature tensor, Ricci tensor and  Ricci
scalar for the metric $g_{ab}$  are related
to those of the metric $\bar{g}_{ab} = e^{2U} g_{ab}$ by formulae given in de Felice and Clarke \cite{felcla}.
The conformal Einstein tensor ${\bf \bar G}$ is given by
\begin{equation}
\bar{G}_{ab} = G_{ab} + 2\left(U_aU_b - \frac{1}{2} U^cU_c g_{ab}\right) +
2(U^c{}_{;c} + U^cU_c)g_{ab} - 2U_{a;b}\;,\label{1f}
\end{equation}
where the covariant derivatives and contractions are calculated on the original
metric $g_{ab}$. An important characteristic of conformal mappings is that the Weyl tensor $C$ remains invariant under the transformation, i.e.
$
\bar{C}_{abcd} = C_{abcd}
$.
A necessary and sufficient condition that a spacetime is conformally
flat is that the Weyl tensor $\bf{C}$ vanishes.
Under a conformal transformation $\bar{g}_{ab} = e^{2U} g_{ab}$,   the velocity field transforms as $
\bar{u}_a = e^Uu_a
$
and we obtain $\bar{\dot u}_a = e^{U} \left( \dot{u}_a + u_au_b U^{,b} + U_{,a}\right)$, $\bar{\Theta} = e^{-U} \Theta - 3u^a \left( e^{-U} \right)_{,a}, $ $\bar{\omega}_{ab} = e^{U} \omega_{ab},  $ $\bar{\sigma}_{ab} = e^{U} \sigma_{ab}$
 for the transformed kinematical quantities acceleration, expansion, vorticity and shear respectively. The above
quantities  will be useful in studying the physical behaviour of the
models generated by a conformal transformation especially if the model is non--static. Barrett and Clarkson \cite{barcla} have calculated how the redshift changes under a conformal transformation. In particular, a formula is provided to compute the redshifts and geodesics of the seed metric.

\section{Vacuum Kasner Spacetime}

In a paper entitled  ``Will the real Kasner metric please stand up?'' A Harvey \cite{har} provided the form
\be
ds^2 = et^{2a_1} dt^2 - et^{2a_2}dx^2 -t^{2a_3}dy^2 - t^{2a_4}dz^2  \label{80}
\ee
for the Kasner metric in cartesian coordinates $(t,x,y,z)$.  The constants $a_k$ satisfy $a_2 +a_3 +a_4 = a_1 +1$, $a_2^2 + a_3^2 + a_4^2 = (a_1 +1)^2$ and $e=\pm 1$. The special case of (\ref{80}) is  the Kasner spacetime  in the more familiar form
\begin{equation}
ds^2 = - dt^2 +t^{2a} dx^2 +t^{2b} dy^2 + t^{2c} dz^2 \label{81}
\end{equation}
 where $a, b$ and $c$ are real numbers satisfying
 \be
 a+b+c=1 = a^2 + b^2 + c^2 \label{81a}
 \ee
 The metric ({\ref{81}) is a plane symmetric non-static solution of the vacuum Einstein field equations. It is homogeneous (position independent) and anisotropic. Contraction of the Kasner universe is guaranteed in at least one of the directions which is evident from the Kasner conditions that force one of $a$, $b$ or $c$ to be negative.

The Riemann tensor components for the spacetime (\ref{81}) have the form
\beq
\frac{t^{2(1-a)}}{a(1-a)} R_{txtx} = \frac{t^{2(1-b)}}{b(1-b)} R_{tyty} =\frac{t^{2(1-c)}}{c(1-c)} R_{tztz} &=& 1 \n \\ \n \\
c t^{2c} R_{xyxy} = bt^{2b} R_{xzxz} = at^{2a} R_{yzyz} &=& abc \n
\eeq
and it is clear that the Kasner metric is not flat.
The Weyl tensor components  are given by
\beq
t^{2(1-a)} C_{txtx} = -t^{2a}C_{yzyz} &=& a(1-a) \n \\ \n \\
t^{2(1-b)} C_{tyty} = -t^{2b} C_{xzxz} &=& b(1-b) \n \\ \n \\
-t^{2(1-c)}C_{tztz} = t^{2c}C_{xyxy} &=& c(1-c) \n
\eeq
and we can note that the Kasner metric is in general not conformally flat. This means that it cannot be expressed as a  multiple of the Minkowski metric.
 The Kretschmann scalar has the following possible values
 \[
 R^{abcd}R_{abcd} = \frac{4\left(3 a^4-12 a^3+20 a^2-16 a+12 \right)}{t^4} \hspace{1cm} \mbox{or} \hspace{1cm} \frac{4 a^2 \left(3 a^2+4\right) }{t^4}
 \]
after invoking (\ref{81a}). This indicates that the metric has an irremovable curvature singularity at $t=0$.
We analyse the conformal counterpart to (\ref{81}) which has the form
\begin{equation}
ds^2 = e^{2U(t,x)} \left(- dt^2 +t^{2a} dx^2 +t^{2b} dx^2 + t^{2c} dz^2 \right) \label{102}
\end{equation}
where the conformal factor $U$ is assumed to be a function of $t$ and the space coordinate  $x$. It can easily be shown that if the conformal factor is of the form $U(x)$ only then the field equations force the homothety $U =$ constant. On the other hand if the form $U=U(t)$ is utilised then the relationships (\ref{81a}) will be violated.   Our intention is to generate perfect fluid models using the seed geometry (\ref{81}).

The Einstein tensor $\tilde{G}^a_b$ is given by
\begin{eqnarray}
\tilde{G}^t_t &=& e^{-2U}\left[\frac{2}{t^{2a}}\left(U_{xx} + U_x^2\right) - 6U_t^2 - \frac{2}{t}U_t\right] \label{5a} \\ \n \\
\tilde{G}^x_t &=&  t^{-2a} \tilde{G}^t_x = 2e^{-2U}\left[U_{tx} -U_tU_x -\frac{a}{t}U_x\right]  \label{5b}  \\ \n \\
\tilde{G}^x_x &=& e^{-2U}\left[-2\left(U_{tt} + U_t^2 + \frac{(b+c)}{t}U_t\right) -\frac{6}{t^{2a}}U_x^2\right] \label{5c} \\ \n \\
\tilde{G}^y_y &=& e^{-2U}\left[ \frac{2}{t^{2a}} \left(U_{xx} + U_x^2\right) - 2\left(U_{tt} + U_t^2 -\frac{(a+c)}{t} U_t\right)\right]   \label{5d} \\ \n \\
\tilde{G}^{z}_{z} &=&  e^{-2U}\left[ \frac{2}{t^{2a}} \left(U_{xx} + U_x^2\right) - 2\left(U_{tt} + U_t^2 -\frac{(a+b)}{t} U_t\right)\right] \label{5e}
\end{eqnarray}
Note we have repeatedly used (\ref{81a}) and elementary identities such as $ac = \frac{1}{2}(a+c)^2 -\frac{1}{2}(a^2 + c^2)$.

It is well known that the Kasner metric itself is anisotropic, however, we ask whether this applies necessarily to its conformal counterpart. Requiring an isotropic particle pressure leads to
\be
 b = c \hspace{1cm} {\mbox {or}} \hspace{1cm} U_t = 0
 \ee
 by comparing (\ref{5d}) and (\ref{5e}). First let us examine the consequences of $U_t = 0$. This means that the conformal factor has the form $U=U(x)$. However, this in turn implies that $U_x = 0$ by virtue of the vanishing of (\ref{5b}). That is the conformal factor is constant thus giving a trivial homothetic transformation that is not interesting physically.

 Next consider the implications of $b=c$. Using (\ref{81a}), this then implies that $b =c = 0$ and so $a = 1$ or $b=c=\frac{2}{3}$ which gives $a = -\frac{1}{3}$. We proceed to analyse each of these possibilities in turn.

 \subsection{The case $b=c=0 , a=1$}

The metric (\ref{102}) assumes the simple form
\begin{equation}
ds^2 = e^{2U(t,x)} \left(- dt^2 +t^{2} dx^2 + dy^2 +  dz^2 \right) \label{103}
\end{equation}
and only one of the spatial directions is impacted by a function of time in the original metric.
The Einstein field equations are given by
 \begin{eqnarray}
\frac{2}{t^{2}}\left(U_{xx} + U_x^2\right) - 6U_t^2 - \frac{2}{t}U_t &=&-\rho e^{4U} \label{7a} \\ \n \\
U_{tx} -U_tU_x -\frac{1}{t}U_x &=& 0 \label{7b}  \\ \n \\
-2\left(U_{tt} + U_t^2 \right) +\frac{6}{t^{2}}U_x^2 &=& pe^{4U} \label{7c} \\ \n \\
 \frac{2}{t^{2}} \left(U_{xx} + U_x^2\right) - 2\left(U_{tt} + U_t^2 -\frac{1}{t} U_t\right) &=& pe^{4U}   \label{7d}
\end{eqnarray}
The general solution of (\ref{7b}) is given by
\be
e^{U} = -\left(t h(x) + g(t)\right)^{-1} \label{71}
\ee
where $h(x)$ and $g(t)$ are functions of integration. This form immediately rules out the possibility of $U$ being a separable function in general.
Additionally, the isotropy condition (\ref{7c}) = (\ref{7d}) yields the constraint
\be
U_{xx}-2U_x^2 + tU_t = 0 \label{72}
\ee
Substituting  (\ref{71}) into (\ref{72}) we obtain
\begin{equation}
g\left(h + {\dot g} + h'' \right) + t\left[h^2 + h'^2 + h\left({\dot g} + h''\right)\right] = 0  \label{73}
\end{equation}
where $ \dot{} \equiv \frac{\partial}{\partial t}$ and $ ' \equiv \frac{\partial}{\partial x}$.
Now equation (\ref{73}) yields solutions only in certain special cases in view of the coupling of $g(t)$ and $h(x)$ and their derivatives. There are two cases that may lead to viable solutions.

\begin{itemize}

\item{} The presence of the factor $t$ outside the second set of parenthesis motivates the form $g(t)=\alpha t$ for some constant $\alpha$ which turns out to be inconsequential as it disappears. In this case equation (\ref{73}) assumes the form
\be
(\alpha + h) h'' + h'^2 + (h+\alpha)^2 = 0 \label{74}
\ee
for the function $h(x)$. This equation is readily solvable in the form
\be
h(x)= -\alpha \pm \sqrt{c_1 \cos \sqrt{2} x - c_2 \sin \sqrt{2} x} \label{75}
\ee
where $c_1$ and $c_2$ are integration constants.

Taking the positive sign before the square root  in (\ref{75})   we obtain
\[
U(t,x)=\ln\left[-\left(t\sqrt{c_1\cos\sqrt{2}x - c_2\sin\sqrt{2}x}\right)^{-1}\right]
\]
 for the conformal factor. In order to admit real valued solutions, it is required that $t < 0$.
The isotropic pressure  and energy density are given by
\begin{eqnarray}
p&=&-\frac{1}{2}t^2(c_1^2+c_2^2+7(c_1-c_2)(c_1+c_2)\cos2\sqrt{2}x-14c_1c_2\sin2\sqrt{2}x)\label{n2} \\ \n \\
\rho&=&\frac{1}{2}t^2(c_1^2+c_2^2+5(c_2^2-c_1^2)\cos2\sqrt{2}x+10c_1c_2\sin2\sqrt{2}x)\label {n3}
\end{eqnarray}
for this case.

Under the conformal transformation $\bar{g}_{ab} = e^{2U}g_{ab}$, the velocity field transforms as
\[
\bar{u}_a = \frac{1}{t\sqrt{c_1\cos\sqrt{2}x-c_2\sin\sqrt{2}x}}
\]
and the following transformed kinematical quantities result
\begin{eqnarray}
\bar{\dot u}_a &=& \left(\frac{2}{t^2\sqrt{c_1\cos\sqrt{2}x-c_2\sin\sqrt{2}x}} , -\frac{c_2\cos\sqrt{2}x+c_1\sin\sqrt{2}x}{\sqrt{2}t\left(c_1\cos\sqrt{2}x-c_2\sin\sqrt{2}x\right)^\frac{3}{2}} \right.\nonumber \\&&\left., 0 , 0\right)\\
\bar{\Theta} &=& 2\sqrt{c_1\cos\sqrt{2}x-c_2\sin\sqrt{2}x} \\ \nonumber \\
\bar{\omega}_{ab} &=& 0\\ \nonumber  \\
\bar{\sigma}_{ab} &=& \left(0 , -\frac{2}{3\sqrt{c_1\cos\sqrt{2}x-c_2\sin\sqrt{2}x}}, \right.\nonumber \\
&&\left.\frac{1}{3t^2\sqrt{c_1\cos\sqrt{2}x-c_2\sin\sqrt{2}x}} , \frac{1}{3t^2\sqrt{c_1\cos\sqrt{2}x-c_2\sin\sqrt{2}x}}\right)
\end{eqnarray}
These calculations reveal that for this form of the conformal map, the fluid congruences are accelerating, expanding, shearing but irrotational in general.

Now considering the negative sign before the square root in (\ref{75})   generates the conformal factor
\begin{equation}
 U(t,x)=\ln\left[\frac{1}{t\sqrt{c_1\cos\sqrt{2}x - c_2\sin\sqrt{2}x})}\right]\label{n1}
 \end{equation}
 and the dynamical variables are given by
\begin{eqnarray}
p&=&-\frac{1}{2}t^2(c_1^2+c_2^2+7(c_1-c_2)(c_1+c_2)\cos2\sqrt{2}x -14c_1c_2\sin2\sqrt{2}x)\label{n2} \\ \n \\
\rho&=&\frac{1}{2}t^2(c_1^2+c_2^2+5(c_2^2-c_1^2)\cos2\sqrt{2}x +10c_1c_2\sin2\sqrt{2}x)\label {n3}
\end{eqnarray}
and the transformed kinematical quantities are as follows :
\begin{eqnarray}
\bar{\dot u}_a &=& \left(-\frac{2}{t^2\sqrt{c_1\cos\sqrt{2}x-c_2\sin\sqrt{2}x}} , \right. \nonumber \\&&\left. \frac{c_2\cos\sqrt{2}x+c_1\sin\sqrt{2}x}{\sqrt{2}t\left(c_1\cos\sqrt{2}x-c_2\sin\sqrt{2}x\right)^\frac{3}{2}} , 0 , 0\right) \\
\bar{\Theta} &=& -2\sqrt{c_1\cos\sqrt{2}x-c_2\sin\sqrt{2}x}
\\
\bar{\omega}_{ab} &=& 0
\\
\bar{\sigma}_{ab} &=& \left(0 , \frac{2}{3\sqrt{c_1\cos\sqrt{2}x-c_2\sin\sqrt{2}x}}, -\frac{1}{3t^2\sqrt{c_1\cos\sqrt{2}x-c_2\sin\sqrt{2}x}} , \right. \nonumber \\&&\left. -\frac{1}{3t^2\sqrt{c_1\cos\sqrt{2}x-c_2\sin\sqrt{2}x}}\right)
\end{eqnarray}
In this case, the perfect fluid is accelerating, collapsing, shearing and non-rotating.

The metric (\ref{103}) may now be written explicitly as
\be
ds^2 = \left[\frac{1}{t^2(c_1\cos\sqrt{2}x - c_2\sin\sqrt{2}x)}\right](- dt^2 +t^2dx^2 + dy^2 + dz^2)\label{n4} \\ \n \\
\ee

The speed of sound in a nonstatic perfect fluid was discussed by Knutsen \cite{knuts,cahill} and was shown to be given by $v^2 = \frac{dp/dt}{d\rho /dt}$ for an adiabatic system with a constant entropy. This followed from the deduction of Nariai \cite{nariai}  that adiabatic flows are characterised by entropy functions that are time independent. Accordingly we obtain
\be
\frac{dp}{d\rho} = - \frac{c_1^2 + c_2^2 + 7(c_1^2 - c_2^2)\cos2\sqrt{2}x - 14c_1c_2\sin2\sqrt{2}x}{c_1^2 + c_2^2 + 5(c_2^2-c_1^2)\cos2\sqrt{2}x + 10c_1c_2\sin2\sqrt{2}x}
\ee
as the sound speed index. It is demanded that $0 < \frac{dp}{d\rho} < 1$ to ensure a subluminal sound speed.

\begin{figure}[t]
\centering
\includegraphics[scale=.8]{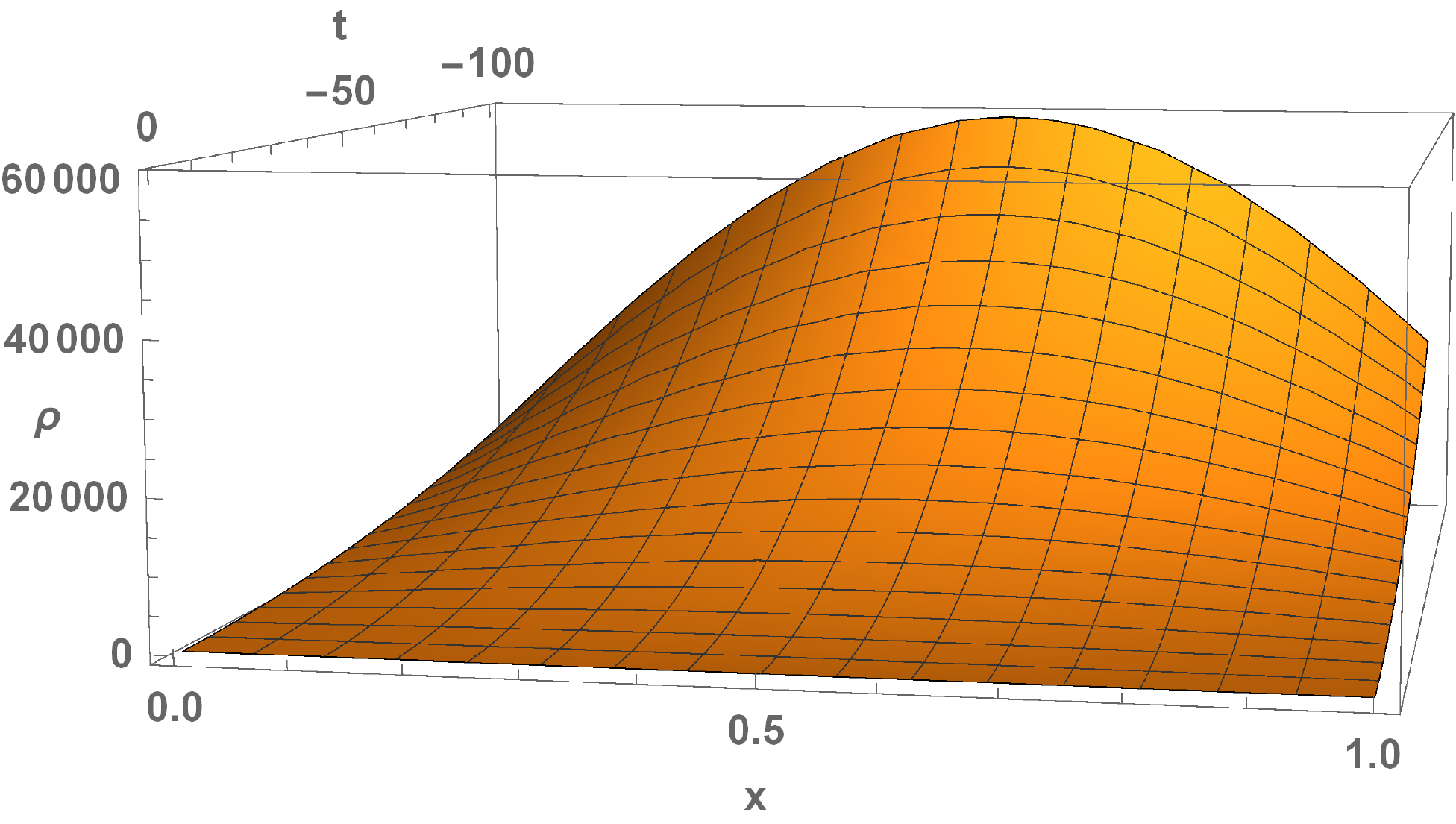}\caption{Energy density as a function of  $x$ and  $t$} \label{rho}
\end{figure}

\begin{figure}[t]
\centering
\includegraphics[scale=.8]{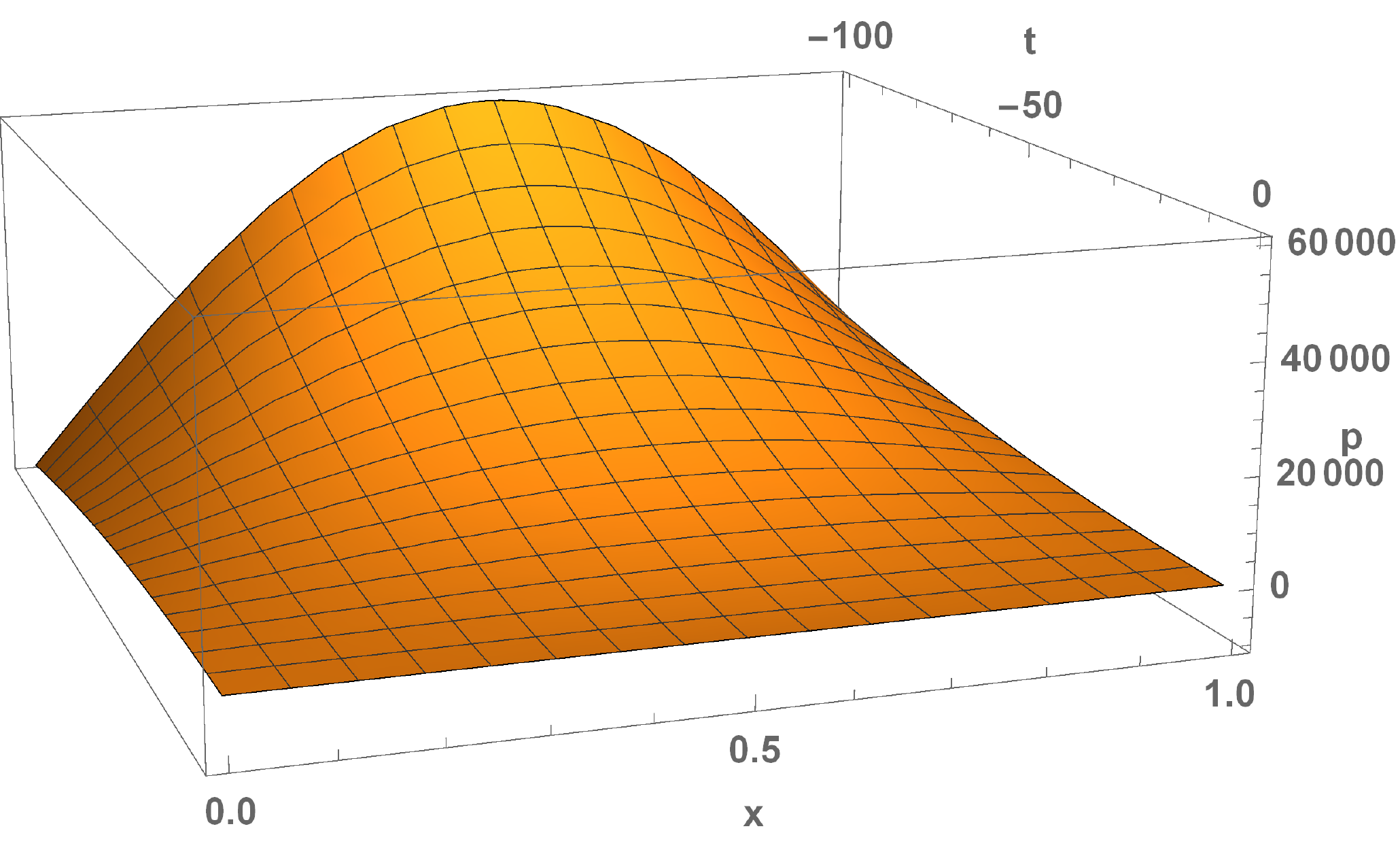}\caption{Pressure as a function of  $x$ and  $t$} \label{p}
\end{figure}

\begin{figure}[t]
\centering
\includegraphics[scale=.8]{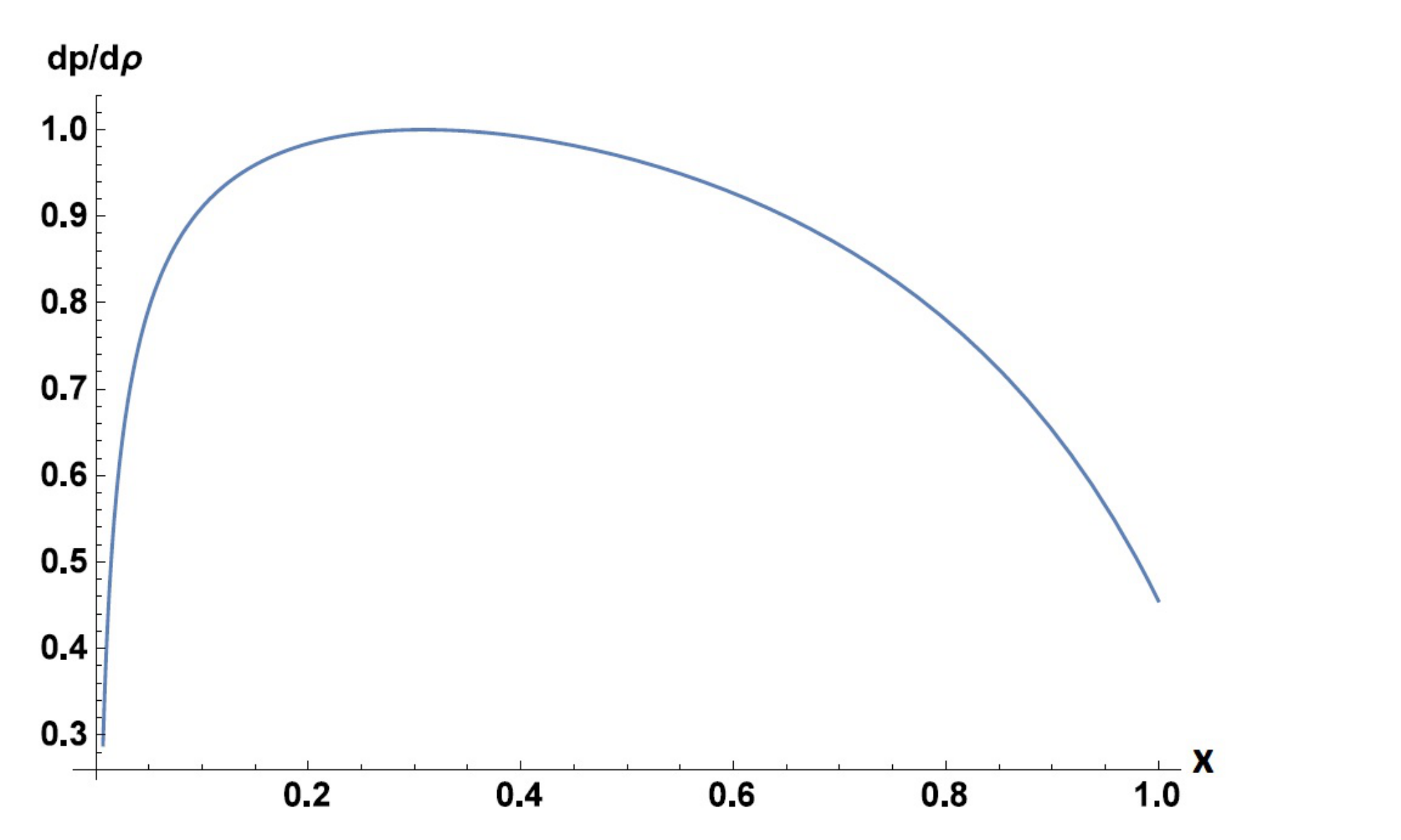}{\caption{Sound speed $dp/d\rho$ as a function of $x$}} \label{dpdrho}
\end{figure}

An examination of the associated plots reveals that the model constructed displays pleasing physical behavior. Fig. 1 and Fig. 2 reflect a density and pressure profile that is smooth and  singularity-free. Fig. 3 depicts the sound speed index $\frac{dp}{d\rho}$ and it may be noted that the requirement $0<\frac{dp}{d\rho} <1$ is satisfied everywhere. This ensures that the speed of sound never exceeds the speed of light in this fluid. It is interesting to observe that the matter variables are well-behaved even though we did not specify an equation of state between the energy density and pressure.

\item{} Equation (\ref{73}) may also be solved with the form $h(x) =$ a constant say $K$. In this case we obtain
\be
(K + \dot{g})(g + Kt)=0
\ee
which is satisfied by $g(t) = -Kt + L$ for some new constant $L$ or just $g(t)= -Kt$ for the second bracket. Plugging this back into the conformal factor reveals that the conformal factor $U$ must be a constant thus yielding a homothety which does not lead to any new physics.

\end{itemize}

\subsection{The case $b=c=\frac{2}{3}$, $a =-\frac{1}{3}$}

For this combination of exponents, the line element (\ref{102}) assumes the form
\begin{equation}
ds^2 = e^{2U(t,x)} \left(- dt^2 +t^{-\frac{2}{3}} dx^2 +t^{\frac{4}{3}} dy^2 + t^{\frac{4}{3}} dz^2 \right) \label{104}
\end{equation}

The Einstein field equations are written as
\begin{eqnarray}
2t^{\frac{2}{3}}\left(U_{xx} + U_x^2\right) - 6U_t^2 - \frac{2}{t}U_t &=& -\rho e^{4U} \label{8a} \\ \n \\
U_{tx} -U_tU_x +\frac{1}{3t}U_x &=& 0  \label{8b}  \\ \n \\
-2\left(U_{tt} + U_t^2 + \frac{4}{3t}U_t\right) - 6t^{\frac{2}{3}} U_x^2 &=& pe^{4U} \label{8c} \\ \n \\
2t^{\frac{2}{3}} \left(U_{xx} + U_x^2\right) - 2\left(U_{tt} + U_t^2 -\frac{4}{3t} U_t\right) &=& pe^{4U}  \label{8d}
\end{eqnarray}

Solving (\ref{8b}) gives the functional form
\begin{equation}\label{n1}
e^U = \left(g(t) - t^{-\frac{1}{3}} \int{ k(x) dx}\right)^{-1}
\end{equation}
which may simply be expressed as
\begin{equation}\label{n2}
e^U = \left(g(t) - t^{-\frac{1}{3}} h(x)\right)^{-1}
\end{equation}
where we have put $h(x) = \int{ k(x) dx}$

 Then equating (\ref{8c}) and (\ref{8d}) to get the isotropy condition yields the equation
\begin{equation}\label{n3}
 U_{xx} + 4U_x^2 + \frac{8}{3}t^{-\frac{5}{3}}U_t = 0
\end{equation}
 Inserting (\ref{n2}) into (\ref{n3}) yields the equation
\begin{equation}
9t^{\frac{8}{3}}\left[h''\left(h - gt^{\frac{1}{3}}\right) + 3h'^2\right] - \left(8h + 24t^{\frac{4}{3}}{\dot g}\right)\left(h - gt^{\frac{1}{3}}\right) = 0 \label{9a}
\end{equation}
Again, the functions $h(x)$ and $g(t)$ are inseparable so special cases could be attempted. Setting $h(x) = q$ a constant, yields $g(t)= q t^{-1/3} + s$  where $s$ is an integration constant. This form results in a constant conformal factor and so is not examined any further. There does not appear to exist any further solutions to (\ref{9a}).

\section{Discussion}

We have utilised the time dependent but anisotropic plane symmetric Kasner spacetime as a seed solution to generate a perfect fluid model with isotropic particle pressure through a conformal mapping. Two categories of solutions were found. Both classes were  non-rotating, accelerating and shearing, however it turned out that one model was expanding while the other was in collapse.   The models generated possessed pleasing physical properties which are consistent with a cosmological fluid. At the basic level the surfaces of density and pressure were smooth, well-behaved and singularity free.  The speed of sound was found not to exceed the speed of light. We have thus succeeded in constructing a viable cosmological fluid distribution with pressure isotropy. This latter condition proved restrictive in solving the field equations. If this is relaxed, then new classes of anisotropic models may be found. This is part of future research. Singularity-free cosmological models have been investigated previously by various authors under the assumption of cylindrical symmetry \cite{patel} and spherical symmetry \cite{dad1}. It is interesting to note that the singularity-free spherical models are necessarily inhomogeneous with heat flow.  Dadhich and Raychaudhuri \cite{dad2} presented an oscillatory inhomogeneous cosmological model with spherical symmetry. The source of matter for this model consists of anisotropic pressure and heat flow. An interesting phenomenon predicted by this model is the possibility of observing blueshifts as the model oscillates between two stable states.

\bibliography{basename of .bib file}


\end{document}